\documentclass[aps,prl,twocolumn,nofootinbib]{revtex4-1} %nofootinbib
\usepackage{epsfig}
\usepackage{color}
\usepackage[latin1]{inputenc}
\usepackage{float,amsmath}
\usepackage{graphicx}
\usepackage{comment}
\usepackage{ulem}

\begin{document}

%\title{System size and flow in p+p, p+Pb and Pb+Pb collisions}
%\title{Proton shape and flow in proton-proton, proton-lead and lead-lead collisions}
\title{Eccentric protons? 
Sensitivity of flow to system size and shape\\ in p+p, p+Pb and Pb+Pb collisions}
\author{Bj\"orn Schenke}
\affiliation{Physics Department, Brookhaven National Laboratory, Upton, NY 11973, USA}
\author{Raju Venugopalan}
\affiliation{Physics Department, Brookhaven National Laboratory, Upton, NY 11973, USA}

\begin{abstract}
We determine the transverse system size of the initial non-equilibrium Glasma state and of the hydrodynamically evolving fireball as a function of produced charged particles in p+p, p+Pb and Pb+Pb collisions at the Large Hadron Collider. 
Our results are consistent with recent measurements of Hanbury-Brown-Twiss (HBT) radii by the ALICE collaboration. 
Azimuthal anisotropy coefficients $v_n$ generated by combining the early time Glasma dynamics with viscous fluid dynamics in Pb+Pb collisions are in excellent agreement with experimental data for a wide range of centralities. In particular, event-by-event distributions of the $v_n$ agree with the experimental data out to fairly peripheral centrality bins.
In striking contrast, our results for p+Pb collisions significantly underestimate the magnitude and do not reproduce the centrality dependence of data for $v_2$ and $v_3$ coefficients. We argue that the measured $v_n$ data and HBT radii strongly constrain the shapes of initial parton distributions across system sizes that would be compatible with a flow interpretation in p+Pb collisions. Alternately, additional sources of correlations may be required to describe the systematics of long range rapidity correlations in p+p and p+Pb collisions.

%While Pb+Pb results are in good agreement with the experimental data, 
%results in p+Pb underestimate the experimental data and do not reproduce the centrality dependence. We argue that an additional source of correlations is needed and/or the description of the proton shape and its fluctuations needs to be modeled differently. 
\end{abstract}
  
%\pacs{11.15Bt, 04.25.Nx, 11.10Wx, 12.38Mh}
\maketitle

%%%%%%%%%%%%%%%%%%%%%%%%%%%%%%%%%%%%%%%%%%%%%%%%%%%%%%%%%%%%

The description of ultra-relativistic heavy-ion (A+A) collisions with event-by-event viscous fluid-dynamic models has been extremely successful \cite{Gale:2013da}. In particular the color-glass-condensate (CGC) \cite{Gelis:2010nm} based IP-Glasma model \cite{Schenke:2012wb,Schenke:2012hg} in combination with the viscous fluid dynamic simulation \textsc{music} \cite{Schenke:2010nt,Schenke:2010rr,Schenke:2011bn} 
%has been able to 
provides a consistent description of particle spectra, the anisotropic flow coefficients $v_n$ and their event-by-event distributions \cite{Gale:2012rq}.

Recent measurements at the Relativistic Heavy Ion Collider (RHIC) at Brookhaven National Laboratory (BNL) and the Large Hadron Collider (LHC) at CERN have shown striking similarities in the structure of long range pseudo-rapidity correlations between high-multiplicity deuteron-gold (d+Au) \cite{Adare:2013piz} and proton-lead (p+Pb) collisions \cite{CMS:2012qk,Abelev:2012ola,Aad:2012gla} and peripheral heavy-ion collisions with similar multiplicity. One may thus conclude that the small collision systems are dominated by the same physics, namely collective flow of the produced matter. Indeed, first fluid dynamic calculations have been able to describe certain features of the experimental data in d+Au and p+Pb collisions \cite{Bozek:2011if,Bozek:2012gr,Bozek:2013df,Bozek:2013uha}. In particular, the observed mass splitting of elliptic flow has been at least qualitatively explained within the fluid dynamic framework \cite{Bozek:2013ska,Werner:2013ipa}.

The observed long range correlations in pseudo-rapidity are an input in the fluid dynamic framework while the azimuthal structure follows from the system's collective response to the transverse geometry as in A+A collisions. An explanation of the long range correlations in all collision systems 
%that does not require the assumption of equal geometries at all rapidities 
is given in the color-glass-condensate based description of multi-particle production in high energy nuclear collisions \cite{Gelis:2008sz,Dusling:2009ni}. In addition, this description produces a collimation in azimuth that is compatible with experimental data on the associated yield in p+p and p+A collisions, without any final state interactions~\cite{Dumitru:2010iy,Dusling:2012iga,Dusling:2012cg,Dusling:2012wy,Dusling:2013oia}.

The important question that needs to be answered is whether the physics responsible for the observed anisotropic flow in A+A collisions is qualitatively
different from that in high-multiplicity p+p and p+A (d+A) collisions, or whether collective effects are always dominant. We argue that in order to conclude the latter, a systematic quantitative description from central to peripheral A+A to p/d+A to p+p collisions needs to be given within the same theoretical framework.

In this letter, we first demonstrate that the IP-Glasma+\textsc{music} model that provided an excellent description of data for central and mid-central A+A collisions at RHIC and LHC continues to provide a good description of the data as we study more and more peripheral heavy-ion events. This holds not only for the mean values of $v_n$ but also their event-by-event distributions. These results are an important validation of the applicability of our model to A+A collisions especially since a recent study concludes that the $v_n$ distributions are not well described by most other initial state models \cite{Renk:2014jja}. We further demonstrate that the system sizes predicted in the IP-Glasma+\textsc{music} model for p+p, p+A, and A+A collisions are compatible with the experimentally measured Hanbury-Brown-Twiss (HBT) radii \cite{Abelev:2014pja}. It is however not possible to determine from the radii alone whether fluid dynamic expansion is present in p+p or p+A collisions. 

We study finally the multiplicity dependence of elliptic and triangular flow in A+A and p+A collisions. This requires a proper description of the multiplicity distribution for both systems \cite{Schenke:2013dpa,Schenke:2012hg}. We find that while the description of $v_2$ and $v_3$ in peripheral A+A collisions is fairly good, the theoretical results for p+A collisions underpredict the experimental data by factors of up to 4. 

Given the excellent results of the model for A+A collisions and the various system sizes, the result for p+Pb collisions has dramatic implications. Two equally exciting explanations for the disagreement are possible. Previously discussed multi-particle correlations present in initial gluon production have been ignored in this and all other calculations that are based on collective final state effects. One explanation of our p+Pb results is that these initial state contributions could significantly modify the result for $v_2$ and $v_3$ if final state effects are not able to overpower them--the latter seems to be the case in A+A collisions \cite{Dusling:2012iga}. Alternatively, the disagreement with the measured $v_2$ and $v_3$ could stem from simplified assumptions about the (spherical) shape of gluon distributions in the proton\footnote{Gluon distributions in the proton are extracted from fits of model parameters to combined H1 and ZEUS data on inclusive structure functions. These give excellent 
$\chi$-squared fits to diffractive and exclusive HERA data~\cite{Rezaeian:2012ji}. However, these data may not fully capture the shapes of gluon distributions.}. Deformed parton distributions in the proton would lead to larger initial eccentricities within our model and could generate significantly larger anisotropic flow. This implies that the new measurements at RHIC and LHC could provide unprecedented insight into the detailed shape of a proton at high energy  \cite{Miller:2008sq,Bjorken:2013boa}. We shall later comment on open questions that both these explanations will have to address.

We begin our systematic study by demonstrating that, for a fixed shear viscosity to entropy density ratio $\eta/s=0.18$, anisotropic flow data from heavy-ion collisions at LHC is well described by fluid dynamic simulations using the IP-Glasma initial state described in \cite{Schenke:2013dpa}. The IP-Glasma energy density and flow velocities serve as input to the fluid dynamic simulation \textsc{music} as described in \cite{Gale:2012rq}. Here we choose the initial time $\tau_0=0.4\,{\rm fm}/c$ for the fluid dynamic simulation.\footnote{The effects of varying $\tau_0$ have been studied previously. They are small even if $\tau_0$ is decreased by a factor of two~\cite{Gale:2012rq}. Increasing $\tau_0$ beyond the quoted value will, in this framework, impact the amount of flow generated.}
We select centralities based on the gluon multiplicity distribution at $\tau_0$, obtained from $\sim 40,000$ IP-Glasma events. This centrality selection method neglects possible corrections due to entropy production during the fluid dynamic evolution and effects from hadronization. It is however close to the experimental procedure and avoids having to simulate the fluid dynamic evolution for tens of thousands of events.
After kinetic freeze-out at $T_{\rm kin~ fo} = 135\,{\rm MeV}$ (chemical freeze-out occurs at $T_{\rm chem~ fo} = 150\,{\rm MeV}$) 
and resonance decays, we determine $v_n$ for $n\in \{2,3,4,5 \}$ of charged hadrons in every event by first determining the event-plane angle 
$ \psi_n=(1/n)\arctan( \langle \sin(n\phi)\rangle/\langle \cos(n\phi)\rangle)\,,$
and then computing $v_n=\langle \cos(n(\phi-\psi_n)) \rangle\,,$
where $\langle \cdot \rangle$ are averages over the charged hadron distribution functions.

\begin{figure}[h]
\includegraphics[width=0.48\textwidth]{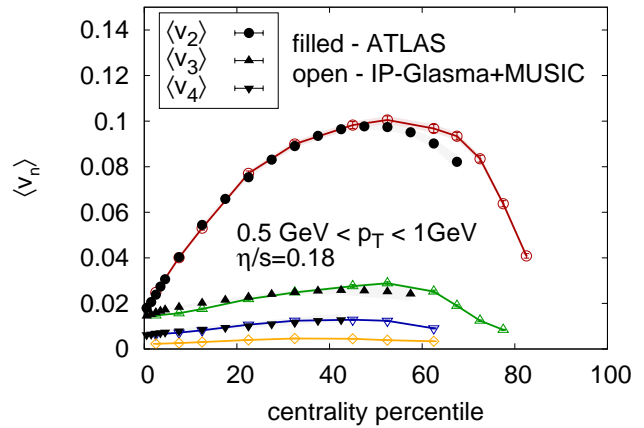}\\
\caption{(Color online) The event averaged $p_T$-integrated $\langle v_n \rangle$ as a function of centrality compared to experimental data from the ATLAS collaboration \cite{Aad:2013xma}. \label{fig:vnCentMean-gtr0-cut}}
\end{figure}

In Fig.\,\ref{fig:vnCentMean-gtr0-cut} we present results for the mean $\langle v_n\rangle$ as a function of centrality compared to experimental results from the ATLAS collaboration \cite{Aad:2013xma}. Here we study significantly more peripheral events than in previous studies \cite{Gale:2012rq}. The  agreement is excellent from the most central to 50\% central events. For more peripheral events our results are up to 10\% larger than the experimental data, with differences being largest for $v_2$. Between 0\% and 20\%, the calculated $v_3$ slightly underestimates the experimental result. 

We next present the computed event-by-event distributions of $v_2$, $v_3$, and $v_4$ and the corresponding initial state eccentricities defined as 
$\varepsilon_n = \sqrt{\langle r^n \cos(n\phi)\rangle^2+\langle r^n \sin(n\phi)\rangle^2}/\langle r^n \rangle$, 
where $\langle \cdot \rangle$ is the average weighted by the deposited energy density.
We compare to data in the respective maximally peripheral bin measured by the ATLAS collaboration \cite{Aad:2013xma}. All distributions are scaled by their mean value. More central bins have been studied previously in \cite{Gale:2012rq}.

 The $\varepsilon_3$ distribution already provides a good description of the measured $v_3$ distribution, while $\varepsilon_2$ and $\varepsilon_4$ distributions are significantly narrower. However, non-linear effects in the fluid dynamic evolution modify the shape of the distributions such that the calculated $v_n$ distributions agree with the experimental result. This result strongly supports the importance of fluid dynamics in heavy-ion collisions.
 We have checked that the scaled distributions are only weakly dependent of the value of $\eta/s$, as was previously found in \cite{Niemi:2012aj}.

\begin{figure}[h]
\includegraphics[width=0.45\textwidth]{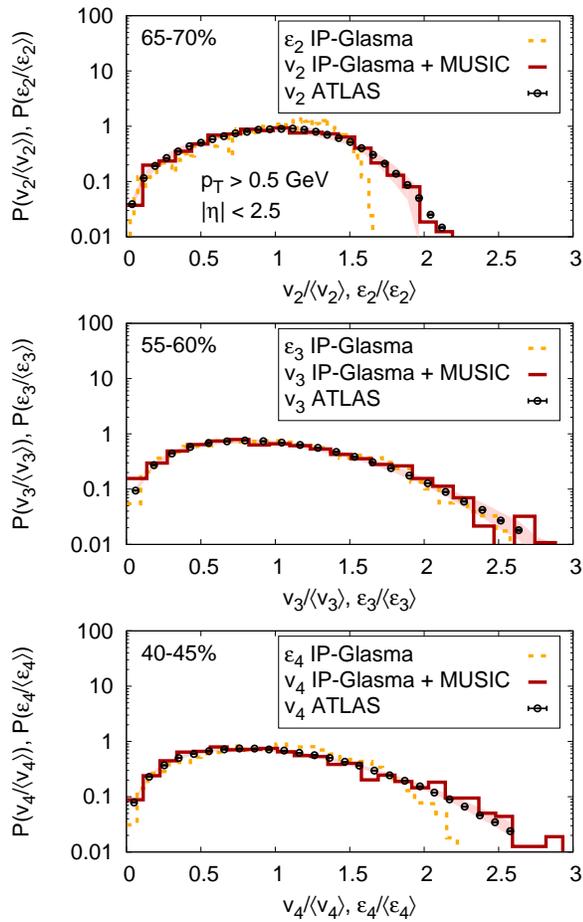}
\caption{(Color online) Data points correspond to the event-by-event distribution of $v_2$, $v_3$, and $v_4$ in the respective maximal peripheral bin measured by the ATLAS collaboration \cite{Aad:2013xma}. These are compared to the distributions of initial eccentricities in the IP-Glasma model and the distributions of $v_n$ from fluid dynamic evolution with IP-Glasma initial conditions. \label{fig:vnenDist-maxCent} }
\end{figure}

Having established that even fairly peripheral events are well described by the IP-Glasma+\textsc{music} model, we now move on to applying the model to p+Pb and p+p collisions. We shall first determine whether the predicted system size (with and without fluid dynamical expansion) is consistent with HBT measurements for all systems.

To be able to compare the initial size as well as the maximal size of the system during the evolution to the measured HBT radii, we define $r_{\rm max}$ as the (angle-averaged) radius where the system reaches the minimal threshold energy density $\varepsilon_{\rm min}$. This defines a size equivalent to the size of the system at freeze-out at a given energy density. This radius by definition depends on the choice of $\varepsilon_{\rm min}$. This choice however only affects the overall normalization of $r_{\rm max}$; it does not affect the dependence of $r_{\rm max}$ on the number of charged particles $N_{\rm ch}$ \cite{Bzdak:2013zma}.
There is also some uncertainty in the radii coming from the choice of the infrared scale $m$ that regulates the long distance tail of the gluon distribution (see \cite{Schenke:2012wb,Schenke:2012hg,Schenke:2013dpa}). 
It can be mostly compensated for by adjusting a normalization constant $K$.

\begin{figure}[h]
\includegraphics[width=0.45\textwidth]{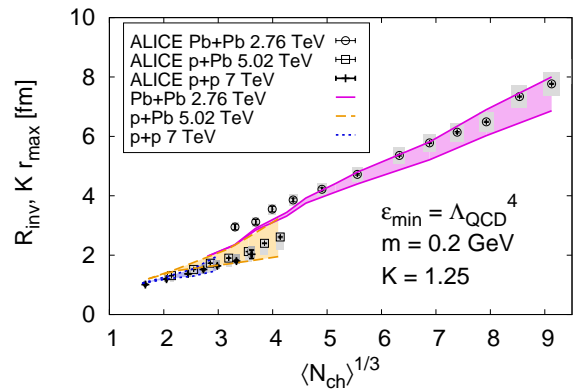}\\
\caption{(Color online) $R_{\rm inv}$ measured by the ALICE collaboration \cite{Abelev:2014pja} compared to $K r_{\rm max}$ determined using the IP-Glasma model and fluid dynamic expansion. The lower end of the band indicates the size of the initial state, the upper end the maximal value of $r_{\rm max}$ during the hydrodynamic evolution. \label{fig:r-Nch-scaled}}
\end{figure}

In Fig.\, \ref{fig:r-Nch-scaled} we show the result for $r_{\rm max}$ in p+p, p+Pb, and Pb+Pb collisions and compare to $R_{\rm inv}$ from the Edgeworth fit to the two-pion correlation function measured by the ALICE collaboration \cite{Abelev:2014pja}. 
We adjust $K$ to match to the p+p results. We determine centrality classes
% in percent 
in the model and assign the $N_{\rm ch}$ value quoted by ALICE \cite{Abelev:2014pja} for each centrality class.

Because the emission of pions occurs throughout the evolution, $R_{\rm inv}$ 
%is expected to 
lies somewhere between the initial radius and the maximal radius reached during evolution. We indicate the range of radii between these two extrema by a band in Fig.\,\ref{fig:r-Nch-scaled}.
We find that our estimate of the system size is compatible with the experimental HBT measurement for all systems simultaneously. 
The Pb+Pb result clearly favors the presence of hydrodynamic expansion.
% the results for p+p and p+Pb show that both the initial state radius and a larger radius within the band are compatible with the data. 

For events with the same multiplicity (for example at $\langle N_{\rm ch}\rangle ^{1/3}\approx 4$),  p+Pb collisions in the hydrodynamic framework show a much more significant expansion compared to Pb+Pb collisions. For these high multiplicities, hydrodynamic expansion in p+Pb collisions appears to be necessary to explain the experimental data.
However, using $m=0.1\,{\rm GeV}$ instead of $m=0.2\,{\rm GeV}$ leads to larger initial radii that are also compatible with the experimental data.

% \begin{figure}[h]
% \includegraphics[width=0.45\textwidth]{r-Nch-scaled-m01}\\
% \caption{(Color online) $R_{\rm inv}$ measured by the ALICE collaboration \cite{Abelev:2014pja} compared to $K r_{\rm max}$ determined using the IP-Glasma model and fluid dynamic expansion. The lower end of the band indicates the size of the initial state, the upper end the maximal value of $r_{\rm max}$ during the hydrodynamic evolution. \label{fig:r-Nch-scaled-m01}}
% \end{figure}

We have established that the details of the bulk properties in Pb+Pb collisions as well as the systematics of the system size from p+p to Pb+Pb collisions are well reproduced in the IP-Glasma (+fluid dynamics) model. We turn now to address anisotropic flow in p+Pb collisions. Using the same method as in Pb+Pb collisions, we determine $v_2$ and $v_3$ as a function of $N_{\rm trk}^{\rm offline}$, measured by the CMS collaboration.\footnote{To obtain $N_{\rm trk}^{\rm offline}$ we determine the centrality class in the IP-Glasma simulations
%in percent 
and match to the $N_{\rm trk}^{\rm offline}$ quoted for that centrality class by the CMS collaboration in \cite{Chatrchyan:2013nka}. $N_{\rm trk}^{\rm offline}\approx 132$ corresponds to 65-70\% central Pb+Pb events, the most peripheral bin shown for the ATLAS data in Fig.\,\ref{fig:vnCentMean-gtr0-cut}.
}

\begin{figure}[h]
\includegraphics[width=0.45\textwidth]{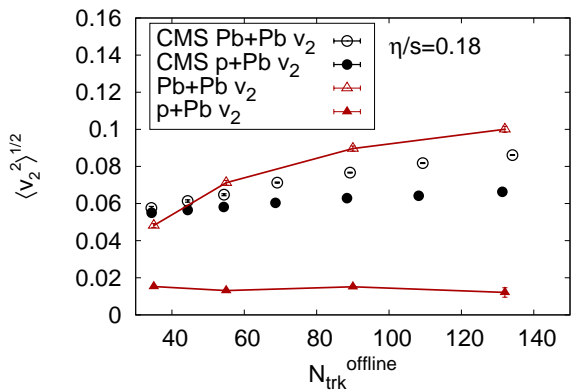}\\
\caption{(Color online) Multiplicity dependence of the root-mean-square elliptic flow coefficient $v_2$ in Pb+Pb (open symbols) and p+Pb collisions (filled symbols) from the IP-Glasma+\textsc{music}
model (connected triangles) compared to experimental data by the CMS collaboration \cite{Chatrchyan:2013nka}.  \label{fig:v2Cent-pPb-CMS}}
\end{figure}

\begin{figure}[h]
\includegraphics[width=0.45\textwidth]{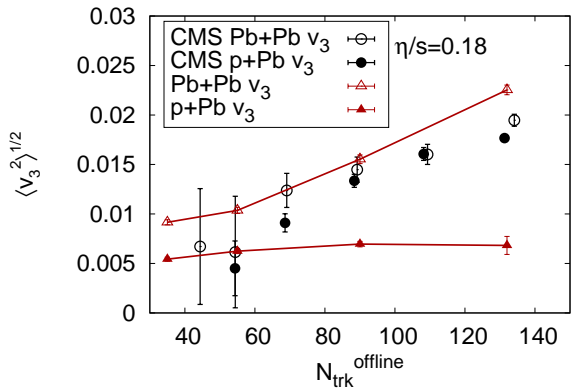}\\
\caption{(Color online) Multiplicity dependence of the root-mean-square triangular flow coefficient $v_3$ in Pb+Pb (open symbols) and p+Pb collisions (filled symbols) from the IP-Glasma+\textsc{music}
model (connected triangles) compared to experimental data by the CMS collaboration \cite{Chatrchyan:2013nka}.   \label{fig:v3Cent-pPb-CMS}}
\end{figure}

Fig.\,\ref{fig:v2Cent-pPb-CMS} shows the calculated $v_2$ in peripheral Pb+Pb collisions and central p+Pb collisions with the same $N_{\rm trk}^{\rm offline}$ in comparison to experimental data by the CMS collaboration \cite{Chatrchyan:2013nka}. 
While the Pb+Pb result reproduces the experimental data within 10-15\%, the computed $v_2$ in p+Pb collisions underestimates the data by a factor of approximately 3.5. We have checked that even in the ideal case ($\eta/s=0$) the data is still underestimated by approximately a factor of 2. We also varied the freeze-out temperature and switching time $\tau_0$, but no choice of parameters could achieve much better agreement with the experimental data. For $v_3$, shown in Fig.\,\ref{fig:v3Cent-pPb-CMS}, we find a similar result: Pb+Pb data are well described, while p+Pb data are underestimated for $N_{\rm trk}^{\rm offline}>60$. Ideal fluid dynamics (not shown) increases the $v_3$ significantly by nearly a factor of 4. Its $N_{\rm trk}^{\rm offline}$ dependence is rather flat, slightly decreasing with increasing $N_{\rm trk}^{\rm offline}$, opposite to the trend seen in the experimental data.

The primary reason for the small $v_n$ in p+Pb collisions is that the initial shape of the system closely follows the shape of the proton (see \cite{Bzdak:2013zma}), which is spherical in our model. The subnucleonic fluctuations included generate non-zero values of the $v_n$, but they do not fully account for the larger experimentally observed values. 
As noted above, modifications of the (fluctuating) proton shape are necessary to account for the larger observed $v_2$ and $v_3$ in p+Pb collisions. If the hydrodynamic paradigm is valid, the results of the high-multiplicity p+Pb and p+p collisions could then in principle be used to extract detailed information on the spatial gluon distribution in the proton. 

There are hydrodynamical models that describe aspects of the p+Pb data. 
These models should also describe key features of Pb+Pb collisions where hydrodynamics is more robust. A model where the spatial geometry of p+Pb collisions is different from ours is that of \cite{Bozek:2011if,Bozek:2012gr,Bozek:2013df,Bozek:2013uha,Bozek:2013ska}, where the interaction region is determined from the geometric positions of participant nucleons. However, as noted, this model falls into the class of models that are claimed \cite{Renk:2014jja} not to be able to reproduce the data on event-by-event $v_n$ distributions in A+A collisions. Whether this particular model can do so needs to be examined. We also note that the $v_2$ centrality dependence in the model differs from the CMS data for p+Pb collisions \cite{Bozek:2013uha}.

Another model which claims large $v_2$ and $v_3$ in p+Pb collisions determines the system size from the position of ``cut pomerons'' and strings \cite{Werner:2013ipa,Werner:2013tya}. The multiplicity dependence of the $v_n$ in this model has not yet been shown. The $v_n$ distributions in A+A collisions should also provide a stringent test of this model.

%Other initial state models, that either determine

%For the former model one would expect much larger system sizes in p+Pb collisions compared to p+p, which would disagree with the experimental data from the ALICE collaboration \cite{Abelev:2014pja}. 
%{\bf problem with Werner's work?}

%mention that Bozek and Werner get better agreement - use different fluctuations (cut pomerons/strings) or MC-Glauber nucleon fluctuations in p+Pb and different energy deposition + NBD \cite{Bozek:2013uha}. mention centrality dependence - Bozek seems to find agreement now (sort of).

In addition to the important quantitative tests imposed on different hydrodynamical models by the experimental data, there 
are conceptual issues that arise due to the possible breakdown of the hydrodynamic  paradigm when extended to very small systems. As shown in recent quantitative studies, viscous corrections can be very significant in p+Pb collisions but play a much smaller role in Pb+Pb collisions \cite{Bzdak:2013zma,Niemi:2014wta}. In particular, an analysis of Knudsen numbers reached during the evolution in A+A and p+A collisions finds that viscous hydrodynamics breaks down for $\eta/s\geq 0.08$ in p+A collisions \cite{Niemi:2014wta}.

An alternative to the hydrodynamic picture and its sensitivity to the proton shape is provided within the Glasma framework itself by initial state correlations of gluons that show a distinct elliptic modulation in relative azimuthal angle \cite{Dumitru:2010iy,Dusling:2012iga,Dusling:2012cg,Dusling:2012wy,Dusling:2013oia}. If these are not overwhelmed in p+Pb collisions by final state effects, as they are in A+A collisions, they can contribute significantly to the observed $v_2$, and possibly $v_3$. The initial state correlations are those of gluons and do not address features of the data such as the mass ordering in particle spectra. While natural in hydrodynamical models, mass ordering may also emerge due to universal hadronization effects, as demonstrated in a string model \cite{Ortiz:2013yxa}.

%I find that correlations from the IP-Glasma are very different when using a common event plane for all p_T or a p_T dependent event plane - data (I think) and hydro do not show this. decorrelation in p_T - how about assoc yield in calculation by Kevin/Raju?

%we checked results for v_n on lattices with 2 lattice spacings that differed by a factor of 2 and found no significant difference
%add other centralities if I have them ...

%mention problem with hydro in small systems (our paper BSTV and Niemi Denicol)

%After having established the validity of the IP-Glasma+\textsc{music} model for peripheral heavy ion collisions, we have shown that the predicted \cite{Bzdak:2013zma} system sizes in p+p, p+Pb, and Pb+Pb collisions are consistent with experimental measurements of HBT radii at the LHC. However, the model fails to reproduce the magnitude and multiplicity dependence of the measured $v_n$ coefficients in p+Pb collisions. 
%This has important implications: It shows that correlation measurements in high-multiplicity p+Pb and p+p collisions are sensitive to the details of the gluon distribution in the protons and nuclei.

In summary, we have shown that the IP-Glasma model in combination with fluid dynamics describes very well the Fourier coefficients of the azimuthal particle distributions in Pb+Pb collisions out to fairly peripheral collisions. Both the experimental mean values and event-by-event distributions are well reproduced. The systematics of HBT radii in p+p, p+A, and A+A collisions are also well described. The discrepancy of our results with the experimental $v_2$ and $v_3$ data in p+Pb collisions therefore poses a challenge to applying the hydrodynamic paradigm to such small systems. A possible solution within the hydrodynamic framework could result from the inclusion of additional shape fluctuations of gluon distributions in both p+p and p+Pb collisions. 
%Then fine tuning might be necessary to describe HBT radii in both p+p and p+Pb collisions.
%Any model incorporating such non-trivial shape fluctuations will also have to reproduce the $v_n$ distributions in A+A  collisions. 
Initial state effects that are also present in the Glasma framework may provide an alternative explanation of the noted discrepancy between experimental data and theory.
These conclusions point to the importance of a deeper understanding of spatial shapes, sizes and correlations of gluon distributions in high energy QCD.

% However, if the fluid dynamic description remains applicable, the disagreement with experimental results in p+Pb collisions shows that these are sensitive to shape fluctuations of the gluon distribution in the proton. This opens up new possibilities to learn about the shape of the proton at high energy and the fluctuations of strong gluon fields from nuclear collisions at RHIC and LHC.

\section*{Acknowledgments}
This research used resources of the National Energy Research Scientific Computing Center, which is supported by the Office of Science of the U.S. Department of Energy under Contract No. DE-AC02-05CH11231. BPS and RV are supported under DOE Contract No. DE-AC02-98CH10886.

%\clearpage
\vspace{-0.5cm}
\bibliography{spires}

\begin{thebibliography}{38}%
\makeatletter
\providecommand \@ifxundefined [1]{%
 \@ifx{#1\undefined}
}%
\providecommand \@ifnum [1]{%
 \ifnum #1\expandafter \@firstoftwo
 \else \expandafter \@secondoftwo
 \fi
}%
\providecommand \@ifx [1]{%
 \ifx #1\expandafter \@firstoftwo
 \else \expandafter \@secondoftwo
 \fi
}%
\providecommand \natexlab [1]{#1}%
\providecommand \enquote  [1]{``#1''}%
\providecommand \bibnamefont  [1]{#1}%
\providecommand \bibfnamefont [1]{#1}%
\providecommand \citenamefont [1]{#1}%
\providecommand \href@noop [0]{\@secondoftwo}%
\providecommand \href [0]{\begingroup \@sanitize@url \@href}%
\providecommand \@href[1]{\@@startlink{#1}\@@href}%
\providecommand \@@href[1]{\endgroup#1\@@endlink}%
\providecommand \@sanitize@url [0]{\catcode `\\12\catcode `\$12\catcode
  `\&12\catcode `\#12\catcode `\^12\catcode `\_12\catcode `\%12\relax}%
\providecommand \@@startlink[1]{}%
\providecommand \@@endlink[0]{}%
\providecommand \url  [0]{\begingroup\@sanitize@url \@url }%
\providecommand \@url [1]{\endgroup\@href {#1}{\urlprefix }}%
\providecommand \urlprefix  [0]{URL }%
\providecommand \Eprint [0]{\href }%
\providecommand \doibase [0]{http://dx.doi.org/}%
\providecommand \selectlanguage [0]{\@gobble}%
\providecommand \bibinfo  [0]{\@secondoftwo}%
\providecommand \bibfield  [0]{\@secondoftwo}%
\providecommand \translation [1]{[#1]}%
\providecommand \BibitemOpen [0]{}%
\providecommand \bibitemStop [0]{}%
\providecommand \bibitemNoStop [0]{.\EOS\space}%
\providecommand \EOS [0]{\spacefactor3000\relax}%
\providecommand \BibitemShut  [1]{\csname bibitem#1\endcsname}%
\let\auto@bib@innerbib\@empty
%</preamble>
\bibitem [{\citenamefont {Gale}\ \emph
  {et~al.}(2013{\natexlab{a}})\citenamefont {Gale}, \citenamefont {Jeon},\ and\
  \citenamefont {Schenke}}]{Gale:2013da}%
  \BibitemOpen
  \bibfield  {author} {\bibinfo {author} {\bibfnamefont {C.}~\bibnamefont
  {Gale}}, \bibinfo {author} {\bibfnamefont {S.}~\bibnamefont {Jeon}}, \ and\
  \bibinfo {author} {\bibfnamefont {B.}~\bibnamefont {Schenke}},\ }\href@noop
  {} {\bibfield  {journal} {\bibinfo  {journal} {Int. J. of Mod. Phys. A, Vol.
  28,}\ }\textbf {\bibinfo {volume} {1340011}} (\bibinfo {year}
  {2013}{\natexlab{a}})},\ \Eprint {http://arxiv.org/abs/1301.5893}
  {arXiv:1301.5893 [nucl-th]} \BibitemShut {NoStop}%
%%CITATION = ARXIV:1301.5893;%%
\bibitem [{\citenamefont {Gelis}\ \emph {et~al.}(2010)\citenamefont {Gelis},
  \citenamefont {Iancu}, \citenamefont {Jalilian-Marian},\ and\ \citenamefont
  {Venugopalan}}]{Gelis:2010nm}%
  \BibitemOpen
  \bibfield  {author} {\bibinfo {author} {\bibfnamefont {F.}~\bibnamefont
  {Gelis}}, \bibinfo {author} {\bibfnamefont {E.}~\bibnamefont {Iancu}},
  \bibinfo {author} {\bibfnamefont {J.}~\bibnamefont {Jalilian-Marian}}, \ and\
  \bibinfo {author} {\bibfnamefont {R.}~\bibnamefont {Venugopalan}},\ }\href
  {\doibase 10.1146/annurev.nucl.010909.083629} {\bibfield  {journal} {\bibinfo
   {journal} {Ann.Rev.Nucl.Part.Sci.}\ }\textbf {\bibinfo {volume} {60}},\
  \bibinfo {pages} {463} (\bibinfo {year} {2010})}\BibitemShut {NoStop}%
\bibitem [{\citenamefont {Schenke}\ \emph
  {et~al.}(2012{\natexlab{a}})\citenamefont {Schenke}, \citenamefont
  {Tribedy},\ and\ \citenamefont {Venugopalan}}]{Schenke:2012wb}%
  \BibitemOpen
  \bibfield  {author} {\bibinfo {author} {\bibfnamefont {B.}~\bibnamefont
  {Schenke}}, \bibinfo {author} {\bibfnamefont {P.}~\bibnamefont {Tribedy}}, \
  and\ \bibinfo {author} {\bibfnamefont {R.}~\bibnamefont {Venugopalan}},\
  }\href@noop {} {\bibfield  {journal} {\bibinfo  {journal} {Phys. Rev. Lett.}\
  }\textbf {\bibinfo {volume} {108}},\ \bibinfo {pages} {252301} (\bibinfo
  {year} {2012}{\natexlab{a}})}\BibitemShut {NoStop}%
%%CITATION = ARXIV:1202.6646;%%
\bibitem [{\citenamefont {Schenke}\ \emph
  {et~al.}(2012{\natexlab{b}})\citenamefont {Schenke}, \citenamefont
  {Tribedy},\ and\ \citenamefont {Venugopalan}}]{Schenke:2012hg}%
  \BibitemOpen
  \bibfield  {author} {\bibinfo {author} {\bibfnamefont {B.}~\bibnamefont
  {Schenke}}, \bibinfo {author} {\bibfnamefont {P.}~\bibnamefont {Tribedy}}, \
  and\ \bibinfo {author} {\bibfnamefont {R.}~\bibnamefont {Venugopalan}},\
  }\href@noop {} {\bibfield  {journal} {\bibinfo  {journal} {Phys. Rev.}\
  }\textbf {\bibinfo {volume} {C86}},\ \bibinfo {pages} {034908} (\bibinfo
  {year} {2012}{\natexlab{b}})}\BibitemShut {NoStop}%
%%CITATION = ARXIV:1206.6805;%%
\bibitem [{\citenamefont {Schenke}\ \emph {et~al.}(2010)\citenamefont
  {Schenke}, \citenamefont {Jeon},\ and\ \citenamefont
  {Gale}}]{Schenke:2010nt}%
  \BibitemOpen
  \bibfield  {author} {\bibinfo {author} {\bibfnamefont {B.}~\bibnamefont
  {Schenke}}, \bibinfo {author} {\bibfnamefont {S.}~\bibnamefont {Jeon}}, \
  and\ \bibinfo {author} {\bibfnamefont {C.}~\bibnamefont {Gale}},\ }\href
  {\doibase 10.1103/PhysRevC.82.014903} {\bibfield  {journal} {\bibinfo
  {journal} {Phys. Rev.}\ }\textbf {\bibinfo {volume} {C82}},\ \bibinfo {pages}
  {014903} (\bibinfo {year} {2010})}\BibitemShut {NoStop}%
%%CITATION = ARXIV:1004.1408;%%
\bibitem [{\citenamefont {Schenke}\ \emph
  {et~al.}(2011{\natexlab{a}})\citenamefont {Schenke}, \citenamefont {Jeon},\
  and\ \citenamefont {Gale}}]{Schenke:2010rr}%
  \BibitemOpen
  \bibfield  {author} {\bibinfo {author} {\bibfnamefont {B.}~\bibnamefont
  {Schenke}}, \bibinfo {author} {\bibfnamefont {S.}~\bibnamefont {Jeon}}, \
  and\ \bibinfo {author} {\bibfnamefont {C.}~\bibnamefont {Gale}},\ }\href
  {\doibase 10.1103/PhysRevLett.106.042301} {\bibfield  {journal} {\bibinfo
  {journal} {Phys. Rev. Lett.}\ }\textbf {\bibinfo {volume} {106}},\ \bibinfo
  {pages} {042301} (\bibinfo {year} {2011}{\natexlab{a}})}\BibitemShut
  {NoStop}%
%%CITATION = 1009.3244;%%
\bibitem [{\citenamefont {Schenke}\ \emph
  {et~al.}(2011{\natexlab{b}})\citenamefont {Schenke}, \citenamefont {Jeon},\
  and\ \citenamefont {Gale}}]{Schenke:2011bn}%
  \BibitemOpen
  \bibfield  {author} {\bibinfo {author} {\bibfnamefont {B.}~\bibnamefont
  {Schenke}}, \bibinfo {author} {\bibfnamefont {S.}~\bibnamefont {Jeon}}, \
  and\ \bibinfo {author} {\bibfnamefont {C.}~\bibnamefont {Gale}},\ }\href
  {\doibase 10.1103/PhysRevC.85.024901} {\bibfield  {journal} {\bibinfo
  {journal} {Phys. Rev.}\ }\textbf {\bibinfo {volume} {C85}},\ \bibinfo {pages}
  {024901} (\bibinfo {year} {2011}{\natexlab{b}})}\BibitemShut {NoStop}%
\bibitem [{\citenamefont {Gale}\ \emph
  {et~al.}(2013{\natexlab{b}})\citenamefont {Gale}, \citenamefont {Jeon},
  \citenamefont {Schenke}, \citenamefont {Tribedy},\ and\ \citenamefont
  {Venugopalan}}]{Gale:2012rq}%
  \BibitemOpen
  \bibfield  {author} {\bibinfo {author} {\bibfnamefont {C.}~\bibnamefont
  {Gale}}, \bibinfo {author} {\bibfnamefont {S.}~\bibnamefont {Jeon}}, \bibinfo
  {author} {\bibfnamefont {B.}~\bibnamefont {Schenke}}, \bibinfo {author}
  {\bibfnamefont {P.}~\bibnamefont {Tribedy}}, \ and\ \bibinfo {author}
  {\bibfnamefont {R.}~\bibnamefont {Venugopalan}},\ }\href {\doibase
  10.1103/PhysRevLett.110.012302} {\bibfield  {journal} {\bibinfo  {journal}
  {Phys.Rev.Lett.}\ }\textbf {\bibinfo {volume} {110}},\ \bibinfo {pages}
  {012302} (\bibinfo {year} {2013}{\natexlab{b}})}\BibitemShut {NoStop}%
%%CITATION = ARXIV:1209.6330;%%
\bibitem [{\citenamefont {Adare}\ \emph {et~al.}(2013)\citenamefont {Adare}
  \emph {et~al.}}]{Adare:2013piz}%
  \BibitemOpen
  \bibfield  {author} {\bibinfo {author} {\bibfnamefont {A.}~\bibnamefont
  {Adare}} \emph {et~al.} (\bibinfo {collaboration} {PHENIX Collaboration}),\
  }\href@noop {} {\  (\bibinfo {year} {2013})},\ \Eprint
  {http://arxiv.org/abs/1303.1794} {arXiv:1303.1794 [nucl-ex]} \BibitemShut
  {NoStop}%
%%CITATION = ARXIV:1303.1794;%%
\bibitem [{\citenamefont {Chatrchyan}\ \emph
  {et~al.}(2013{\natexlab{a}})\citenamefont {Chatrchyan} \emph
  {et~al.}}]{CMS:2012qk}%
  \BibitemOpen
  \bibfield  {author} {\bibinfo {author} {\bibfnamefont {S.}~\bibnamefont
  {Chatrchyan}} \emph {et~al.} (\bibinfo {collaboration} {CMS Collaboration}),\
  }\href {\doibase 10.1016/j.physletb.2012.11.025} {\bibfield  {journal}
  {\bibinfo  {journal} {Phys.Lett.}\ }\textbf {\bibinfo {volume} {B718}},\
  \bibinfo {pages} {795} (\bibinfo {year} {2013}{\natexlab{a}})},\ \bibinfo
  {note} {supplemental materials:
  https://twiki.cern.ch/twiki/bin/view/CMSPublic/ PhysicsResultsHIN12015},\
  \Eprint {http://arxiv.org/abs/1210.5482} {arXiv:1210.5482 [nucl-ex]}
  \BibitemShut {NoStop}%
%%CITATION = ARXIV:1210.5482;%%
\bibitem [{\citenamefont {Abelev}\ \emph {et~al.}(2013)\citenamefont {Abelev}
  \emph {et~al.}}]{Abelev:2012ola}%
  \BibitemOpen
  \bibfield  {author} {\bibinfo {author} {\bibfnamefont {B.}~\bibnamefont
  {Abelev}} \emph {et~al.} (\bibinfo {collaboration} {ALICE Collaboration}),\
  }\href {\doibase 10.1016/j.physletb.2013.01.012} {\bibfield  {journal}
  {\bibinfo  {journal} {Phys.Lett.}\ }\textbf {\bibinfo {volume} {B719}},\
  \bibinfo {pages} {29} (\bibinfo {year} {2013})},\ \Eprint
  {http://arxiv.org/abs/1212.2001} {arXiv:1212.2001 [nucl-ex]} \BibitemShut
  {NoStop}%
%%CITATION = ARXIV:1212.2001;%%
\bibitem [{\citenamefont {Aad}\ \emph {et~al.}(2012)\citenamefont {Aad} \emph
  {et~al.}}]{Aad:2012gla}%
  \BibitemOpen
  \bibfield  {author} {\bibinfo {author} {\bibfnamefont {G.}~\bibnamefont
  {Aad}} \emph {et~al.} (\bibinfo {collaboration} {ATLAS Collaboration}),\
  }\href@noop {} {\  (\bibinfo {year} {2012})},\ \Eprint
  {http://arxiv.org/abs/1212.5198} {arXiv:1212.5198 [hep-ex]} \BibitemShut
  {NoStop}%
%%CITATION = ARXIV:1212.5198;%%
\bibitem [{\citenamefont {Bozek}(2012)}]{Bozek:2011if}%
  \BibitemOpen
  \bibfield  {author} {\bibinfo {author} {\bibfnamefont {P.}~\bibnamefont
  {Bozek}},\ }\href {\doibase 10.1103/PhysRevC.85.014911} {\bibfield  {journal}
  {\bibinfo  {journal} {Phys.Rev.}\ }\textbf {\bibinfo {volume} {C85}},\
  \bibinfo {pages} {014911} (\bibinfo {year} {2012})}\BibitemShut {NoStop}%
%%CITATION = ARXIV:1112.0915;%%
\bibitem [{\citenamefont {Bozek}\ and\ \citenamefont
  {Broniowski}(2013{\natexlab{a}})}]{Bozek:2012gr}%
  \BibitemOpen
  \bibfield  {author} {\bibinfo {author} {\bibfnamefont {P.}~\bibnamefont
  {Bozek}}\ and\ \bibinfo {author} {\bibfnamefont {W.}~\bibnamefont
  {Broniowski}},\ }\href {\doibase 10.1016/j.physletb.2012.12.051} {\bibfield
  {journal} {\bibinfo  {journal} {Phys.Lett.}\ }\textbf {\bibinfo {volume}
  {B718}},\ \bibinfo {pages} {1557} (\bibinfo {year} {2013}{\natexlab{a}})},\
  \Eprint {http://arxiv.org/abs/1211.0845} {arXiv:1211.0845 [nucl-th]}
  \BibitemShut {NoStop}%
%%CITATION = ARXIV:1211.0845;%%
\bibitem [{\citenamefont {Bozek}\ and\ \citenamefont
  {Broniowski}(2013{\natexlab{b}})}]{Bozek:2013df}%
  \BibitemOpen
  \bibfield  {author} {\bibinfo {author} {\bibfnamefont {P.}~\bibnamefont
  {Bozek}}\ and\ \bibinfo {author} {\bibfnamefont {W.}~\bibnamefont
  {Broniowski}},\ }\href@noop {} {\  (\bibinfo {year} {2013}{\natexlab{b}})},\
  \Eprint {http://arxiv.org/abs/1301.3314} {arXiv:1301.3314 [nucl-th]}
  \BibitemShut {NoStop}%
%%CITATION = ARXIV:1301.3314;%%
\bibitem [{\citenamefont {Bozek}\ and\ \citenamefont
  {Broniowski}(2013{\natexlab{c}})}]{Bozek:2013uha}%
  \BibitemOpen
  \bibfield  {author} {\bibinfo {author} {\bibfnamefont {P.}~\bibnamefont
  {Bozek}}\ and\ \bibinfo {author} {\bibfnamefont {W.}~\bibnamefont
  {Broniowski}},\ }\href {\doibase 10.1103/PhysRevC.88.014903} {\bibfield
  {journal} {\bibinfo  {journal} {Phys.Rev.}\ }\textbf {\bibinfo {volume}
  {C88}},\ \bibinfo {pages} {014903} (\bibinfo {year} {2013}{\natexlab{c}})},\
  \Eprint {http://arxiv.org/abs/1304.3044} {arXiv:1304.3044 [nucl-th]}
  \BibitemShut {NoStop}%
%%CITATION = ARXIV:1304.3044;%%
\bibitem [{\citenamefont {Bozek}\ \emph {et~al.}(2013)\citenamefont {Bozek},
  \citenamefont {Broniowski},\ and\ \citenamefont {Torrieri}}]{Bozek:2013ska}%
  \BibitemOpen
  \bibfield  {author} {\bibinfo {author} {\bibfnamefont {P.}~\bibnamefont
  {Bozek}}, \bibinfo {author} {\bibfnamefont {W.}~\bibnamefont {Broniowski}}, \
  and\ \bibinfo {author} {\bibfnamefont {G.}~\bibnamefont {Torrieri}},\ }\href
  {\doibase 10.1103/PhysRevLett.111.172303} {\bibfield  {journal} {\bibinfo
  {journal} {Phys.Rev.Lett.}\ }\textbf {\bibinfo {volume} {111}},\ \bibinfo
  {pages} {172303} (\bibinfo {year} {2013})},\ \Eprint
  {http://arxiv.org/abs/1307.5060} {arXiv:1307.5060 [nucl-th]} \BibitemShut
  {NoStop}%
%%CITATION = ARXIV:1307.5060;%%
\bibitem [{\citenamefont {Werner}\ \emph
  {et~al.}(2013{\natexlab{a}})\citenamefont {Werner}, \citenamefont {Bleicher},
  \citenamefont {Guiot}, \citenamefont {Karpenko},\ and\ \citenamefont
  {Pierog}}]{Werner:2013ipa}%
  \BibitemOpen
  \bibfield  {author} {\bibinfo {author} {\bibfnamefont {K.}~\bibnamefont
  {Werner}}, \bibinfo {author} {\bibfnamefont {M.}~\bibnamefont {Bleicher}},
  \bibinfo {author} {\bibfnamefont {B.}~\bibnamefont {Guiot}}, \bibinfo
  {author} {\bibfnamefont {I.}~\bibnamefont {Karpenko}}, \ and\ \bibinfo
  {author} {\bibfnamefont {T.}~\bibnamefont {Pierog}},\ }\href@noop {} {\
  (\bibinfo {year} {2013}{\natexlab{a}})},\ \Eprint
  {http://arxiv.org/abs/1307.4379} {arXiv:1307.4379} \BibitemShut {NoStop}%
%%CITATION = ARXIV:1307.4379;%%
\bibitem [{\citenamefont {Gelis}\ \emph {et~al.}(2008)\citenamefont {Gelis},
  \citenamefont {Lappi},\ and\ \citenamefont {Venugopalan}}]{Gelis:2008sz}%
  \BibitemOpen
  \bibfield  {author} {\bibinfo {author} {\bibfnamefont {F.}~\bibnamefont
  {Gelis}}, \bibinfo {author} {\bibfnamefont {T.}~\bibnamefont {Lappi}}, \ and\
  \bibinfo {author} {\bibfnamefont {R.}~\bibnamefont {Venugopalan}},\ }\href
  {\doibase 10.1103/PhysRevD.79.094017} {\bibfield  {journal} {\bibinfo
  {journal} {Phys. Rev.}\ }\textbf {\bibinfo {volume} {D79}},\ \bibinfo {pages}
  {094017} (\bibinfo {year} {2008})}\BibitemShut {NoStop}%
%%CITATION = 0810.4829;%%
\bibitem [{\citenamefont {Dusling}\ \emph {et~al.}(2010)\citenamefont
  {Dusling}, \citenamefont {Gelis}, \citenamefont {Lappi},\ and\ \citenamefont
  {Venugopalan}}]{Dusling:2009ni}%
  \BibitemOpen
  \bibfield  {author} {\bibinfo {author} {\bibfnamefont {K.}~\bibnamefont
  {Dusling}}, \bibinfo {author} {\bibfnamefont {F.}~\bibnamefont {Gelis}},
  \bibinfo {author} {\bibfnamefont {T.}~\bibnamefont {Lappi}}, \ and\ \bibinfo
  {author} {\bibfnamefont {R.}~\bibnamefont {Venugopalan}},\ }\href {\doibase
  10.1016/j.nuclphysa.2009.12.044} {\bibfield  {journal} {\bibinfo  {journal}
  {Nucl. Phys.}\ }\textbf {\bibinfo {volume} {A836}},\ \bibinfo {pages} {159}
  (\bibinfo {year} {2010})},\ \Eprint {http://arxiv.org/abs/0911.2720}
  {arXiv:0911.2720 [hep-ph]} \BibitemShut {NoStop}%
%%CITATION = 0911.2720;%%
\bibitem [{\citenamefont {Dumitru}\ \emph {et~al.}(2011)\citenamefont
  {Dumitru}, \citenamefont {Dusling}, \citenamefont {Gelis}, \citenamefont
  {Jalilian-Marian}, \citenamefont {Lappi},\ and\ \citenamefont
  {Venugopalan}}]{Dumitru:2010iy}%
  \BibitemOpen
  \bibfield  {author} {\bibinfo {author} {\bibfnamefont {A.}~\bibnamefont
  {Dumitru}}, \bibinfo {author} {\bibfnamefont {K.}~\bibnamefont {Dusling}},
  \bibinfo {author} {\bibfnamefont {F.}~\bibnamefont {Gelis}}, \bibinfo
  {author} {\bibfnamefont {J.}~\bibnamefont {Jalilian-Marian}}, \bibinfo
  {author} {\bibfnamefont {T.}~\bibnamefont {Lappi}}, \ and\ \bibinfo {author}
  {\bibfnamefont {R.}~\bibnamefont {Venugopalan}},\ }\href {\doibase
  10.1016/j.physletb.2011.01.024} {\bibfield  {journal} {\bibinfo  {journal}
  {Phys. Lett.}\ }\textbf {\bibinfo {volume} {B697}},\ \bibinfo {pages} {21}
  (\bibinfo {year} {2011})},\ \Eprint {http://arxiv.org/abs/1009.5295}
  {arXiv:1009.5295 [hep-ph]} \BibitemShut {NoStop}%
\bibitem [{\citenamefont {Dusling}\ and\ \citenamefont
  {Venugopalan}(2012{\natexlab{a}})}]{Dusling:2012iga}%
  \BibitemOpen
  \bibfield  {author} {\bibinfo {author} {\bibfnamefont {K.}~\bibnamefont
  {Dusling}}\ and\ \bibinfo {author} {\bibfnamefont {R.}~\bibnamefont
  {Venugopalan}},\ }\href {\doibase 10.1103/PhysRevLett.108.262001} {\bibfield
  {journal} {\bibinfo  {journal} {Phys.Rev.Lett.}\ }\textbf {\bibinfo {volume}
  {108}},\ \bibinfo {pages} {262001} (\bibinfo {year} {2012}{\natexlab{a}})},\
  \Eprint {http://arxiv.org/abs/1201.2658} {arXiv:1201.2658 [hep-ph]}
  \BibitemShut {NoStop}%
%%CITATION = ARXIV:1201.2658;%%
\bibitem [{\citenamefont {Dusling}\ and\ \citenamefont
  {Venugopalan}(2012{\natexlab{b}})}]{Dusling:2012cg}%
  \BibitemOpen
  \bibfield  {author} {\bibinfo {author} {\bibfnamefont {K.}~\bibnamefont
  {Dusling}}\ and\ \bibinfo {author} {\bibfnamefont {R.}~\bibnamefont
  {Venugopalan}},\ }\href@noop {} {\  (\bibinfo {year} {2012}{\natexlab{b}})},\
  \Eprint {http://arxiv.org/abs/1210.3890} {arXiv:1210.3890 [hep-ph]}
  \BibitemShut {NoStop}%
%%CITATION = ARXIV:1210.3890;%%
\bibitem [{\citenamefont {Dusling}\ and\ \citenamefont
  {Venugopalan}(2012{\natexlab{c}})}]{Dusling:2012wy}%
  \BibitemOpen
  \bibfield  {author} {\bibinfo {author} {\bibfnamefont {K.}~\bibnamefont
  {Dusling}}\ and\ \bibinfo {author} {\bibfnamefont {R.}~\bibnamefont
  {Venugopalan}},\ }\href@noop {} {\  (\bibinfo {year} {2012}{\natexlab{c}})},\
  \Eprint {http://arxiv.org/abs/1211.3701} {arXiv:1211.3701 [hep-ph]}
  \BibitemShut {NoStop}%
%%CITATION = ARXIV:1211.3701;%%
\bibitem [{\citenamefont {Dusling}\ and\ \citenamefont
  {Venugopalan}(2013)}]{Dusling:2013oia}%
  \BibitemOpen
  \bibfield  {author} {\bibinfo {author} {\bibfnamefont {K.}~\bibnamefont
  {Dusling}}\ and\ \bibinfo {author} {\bibfnamefont {R.}~\bibnamefont
  {Venugopalan}},\ }\href {\doibase 10.1103/PhysRevD.87.094034} {\bibfield
  {journal} {\bibinfo  {journal} {Phys.Rev.}\ }\textbf {\bibinfo {volume}
  {D87}},\ \bibinfo {pages} {094034} (\bibinfo {year} {2013})},\ \Eprint
  {http://arxiv.org/abs/1302.7018} {arXiv:1302.7018 [hep-ph]} \BibitemShut
  {NoStop}%
%%CITATION = ARXIV:1302.7018;%%
\bibitem [{\citenamefont {Renk}\ and\ \citenamefont
  {Niemi}(2014)}]{Renk:2014jja}%
  \BibitemOpen
  \bibfield  {author} {\bibinfo {author} {\bibfnamefont {T.}~\bibnamefont
  {Renk}}\ and\ \bibinfo {author} {\bibfnamefont {H.}~\bibnamefont {Niemi}},\
  }\href@noop {} {\  (\bibinfo {year} {2014})},\ \Eprint
  {http://arxiv.org/abs/1401.2069} {arXiv:1401.2069 [nucl-th]} \BibitemShut
  {NoStop}%
%%CITATION = ARXIV:1401.2069;%%
\bibitem [{\citenamefont {Abelev}\ \emph {et~al.}(2014)\citenamefont {Abelev}
  \emph {et~al.}}]{Abelev:2014pja}%
  \BibitemOpen
  \bibfield  {author} {\bibinfo {author} {\bibfnamefont {B.~B.}\ \bibnamefont
  {Abelev}} \emph {et~al.} (\bibinfo {collaboration} {ALICE Collaboration}),\
  }\href@noop {} {\  (\bibinfo {year} {2014})},\ \Eprint
  {http://arxiv.org/abs/1404.1194} {arXiv:1404.1194 [nucl-ex]} \BibitemShut
  {NoStop}%
%%CITATION = ARXIV:1404.1194;%%
\bibitem [{\citenamefont {Schenke}\ \emph {et~al.}(2013)\citenamefont
  {Schenke}, \citenamefont {Tribedy},\ and\ \citenamefont
  {Venugopalan}}]{Schenke:2013dpa}%
  \BibitemOpen
  \bibfield  {author} {\bibinfo {author} {\bibfnamefont {B.}~\bibnamefont
  {Schenke}}, \bibinfo {author} {\bibfnamefont {P.}~\bibnamefont {Tribedy}}, \
  and\ \bibinfo {author} {\bibfnamefont {R.}~\bibnamefont {Venugopalan}},\
  }\href@noop {} {\  (\bibinfo {year} {2013})},\ \Eprint
  {http://arxiv.org/abs/1311.3636} {arXiv:1311.3636 [hep-ph]} \BibitemShut
  {NoStop}%
%%CITATION = ARXIV:1311.3636;%%
\bibitem [{\citenamefont {Rezaeian}\ \emph {et~al.}(2013)\citenamefont
  {Rezaeian}, \citenamefont {Siddikov}, \citenamefont {Van~de Klundert},\ and\
  \citenamefont {Venugopalan}}]{Rezaeian:2012ji}%
  \BibitemOpen
  \bibfield  {author} {\bibinfo {author} {\bibfnamefont {A.~H.}\ \bibnamefont
  {Rezaeian}}, \bibinfo {author} {\bibfnamefont {M.}~\bibnamefont {Siddikov}},
  \bibinfo {author} {\bibfnamefont {M.}~\bibnamefont {Van~de Klundert}}, \ and\
  \bibinfo {author} {\bibfnamefont {R.}~\bibnamefont {Venugopalan}},\
  }\href@noop {} {\bibfield  {journal} {\bibinfo  {journal} {Phys.Rev.}\
  }\textbf {\bibinfo {volume} {D87}},\ \bibinfo {pages} {034002} (\bibinfo
  {year} {2013})}\BibitemShut {NoStop}%
%%CITATION = ARXIV:1212.2974;%%
\bibitem [{\citenamefont {Miller}(2008)}]{Miller:2008sq}%
  \BibitemOpen
  \bibfield  {author} {\bibinfo {author} {\bibfnamefont {G.~A.}\ \bibnamefont
  {Miller}},\ }\href {\doibase 10.1080/10506890802123721} {\bibfield  {journal}
  {\bibinfo  {journal} {Nucl.Phys.News}\ }\textbf {\bibinfo {volume} {18}},\
  \bibinfo {pages} {12} (\bibinfo {year} {2008})},\ \Eprint
  {http://arxiv.org/abs/0802.3731} {arXiv:0802.3731 [nucl-th]} \BibitemShut
  {NoStop}%
%%CITATION = ARXIV:0802.3731;%%
\bibitem [{\citenamefont {Bjorken}\ \emph {et~al.}(2013)\citenamefont
  {Bjorken}, \citenamefont {Brodsky},\ and\ \citenamefont
  {Scharff~Goldhaber}}]{Bjorken:2013boa}%
  \BibitemOpen
  \bibfield  {author} {\bibinfo {author} {\bibfnamefont {J.~D.}\ \bibnamefont
  {Bjorken}}, \bibinfo {author} {\bibfnamefont {S.~J.}\ \bibnamefont
  {Brodsky}}, \ and\ \bibinfo {author} {\bibfnamefont {A.}~\bibnamefont
  {Scharff~Goldhaber}},\ }\href {\doibase 10.1016/j.physletb.2013.08.066}
  {\bibfield  {journal} {\bibinfo  {journal} {Phys.Lett.}\ }\textbf {\bibinfo
  {volume} {B726}},\ \bibinfo {pages} {344} (\bibinfo {year} {2013})},\ \Eprint
  {http://arxiv.org/abs/1308.1435} {arXiv:1308.1435 [hep-ph]} \BibitemShut
  {NoStop}%
%%CITATION = ARXIV:1308.1435;%%
\bibitem [{\citenamefont {Aad}\ \emph {et~al.}(2013)\citenamefont {Aad} \emph
  {et~al.}}]{Aad:2013xma}%
  \BibitemOpen
  \bibfield  {author} {\bibinfo {author} {\bibfnamefont {G.}~\bibnamefont
  {Aad}} \emph {et~al.} (\bibinfo {collaboration} {ATLAS Collaboration}),\
  }\href {\doibase 10.1007/JHEP11(2013)183} {\bibfield  {journal} {\bibinfo
  {journal} {JHEP}\ }\textbf {\bibinfo {volume} {1311}},\ \bibinfo {pages}
  {183} (\bibinfo {year} {2013})},\ \Eprint {http://arxiv.org/abs/1305.2942}
  {arXiv:1305.2942 [hep-ex]} \BibitemShut {NoStop}%
%%CITATION = ARXIV:1305.2942;%%
\bibitem [{\citenamefont {Niemi}\ \emph {et~al.}(2013)\citenamefont {Niemi},
  \citenamefont {Denicol}, \citenamefont {Holopainen},\ and\ \citenamefont
  {Huovinen}}]{Niemi:2012aj}%
  \BibitemOpen
  \bibfield  {author} {\bibinfo {author} {\bibfnamefont {H.}~\bibnamefont
  {Niemi}}, \bibinfo {author} {\bibfnamefont {G.}~\bibnamefont {Denicol}},
  \bibinfo {author} {\bibfnamefont {H.}~\bibnamefont {Holopainen}}, \ and\
  \bibinfo {author} {\bibfnamefont {P.}~\bibnamefont {Huovinen}},\ }\href
  {\doibase 10.1103/PhysRevC.87.054901} {\bibfield  {journal} {\bibinfo
  {journal} {Phys.Rev.}\ }\textbf {\bibinfo {volume} {C87}},\ \bibinfo {pages}
  {054901} (\bibinfo {year} {2013})},\ \Eprint {http://arxiv.org/abs/1212.1008}
  {arXiv:1212.1008 [nucl-th]} \BibitemShut {NoStop}%
%%CITATION = ARXIV:1212.1008;%%
\bibitem [{\citenamefont {Bzdak}\ \emph {et~al.}(2013)\citenamefont {Bzdak},
  \citenamefont {Schenke}, \citenamefont {Tribedy},\ and\ \citenamefont
  {Venugopalan}}]{Bzdak:2013zma}%
  \BibitemOpen
  \bibfield  {author} {\bibinfo {author} {\bibfnamefont {A.}~\bibnamefont
  {Bzdak}}, \bibinfo {author} {\bibfnamefont {B.}~\bibnamefont {Schenke}},
  \bibinfo {author} {\bibfnamefont {P.}~\bibnamefont {Tribedy}}, \ and\
  \bibinfo {author} {\bibfnamefont {R.}~\bibnamefont {Venugopalan}},\ }\href
  {\doibase 10.1103/PhysRevC.87.064906} {\bibfield  {journal} {\bibinfo
  {journal} {Phys.Rev.}\ }\textbf {\bibinfo {volume} {C87}},\ \bibinfo {pages}
  {064906} (\bibinfo {year} {2013})},\ \Eprint {http://arxiv.org/abs/1304.3403}
  {arXiv:1304.3403 [nucl-th]} \BibitemShut {NoStop}%
%%CITATION = ARXIV:1304.3403;%%
\bibitem [{\citenamefont {Chatrchyan}\ \emph
  {et~al.}(2013{\natexlab{b}})\citenamefont {Chatrchyan} \emph
  {et~al.}}]{Chatrchyan:2013nka}%
  \BibitemOpen
  \bibfield  {author} {\bibinfo {author} {\bibfnamefont {S.}~\bibnamefont
  {Chatrchyan}} \emph {et~al.} (\bibinfo {collaboration} {CMS Collaboration}),\
  }\href {\doibase 10.1016/j.physletb.2013.06.028} {\bibfield  {journal}
  {\bibinfo  {journal} {Phys.Lett.}\ }\textbf {\bibinfo {volume} {B724}},\
  \bibinfo {pages} {213} (\bibinfo {year} {2013}{\natexlab{b}})},\ \Eprint
  {http://arxiv.org/abs/1305.0609} {arXiv:1305.0609 [nucl-ex]} \BibitemShut
  {NoStop}%
%%CITATION = ARXIV:1305.0609;%%
\bibitem [{\citenamefont {Werner}\ \emph
  {et~al.}(2013{\natexlab{b}})\citenamefont {Werner}, \citenamefont {Guiot},
  \citenamefont {Karpenko},\ and\ \citenamefont {Pierog}}]{Werner:2013tya}%
  \BibitemOpen
  \bibfield  {author} {\bibinfo {author} {\bibfnamefont {K.}~\bibnamefont
  {Werner}}, \bibinfo {author} {\bibfnamefont {B.}~\bibnamefont {Guiot}},
  \bibinfo {author} {\bibfnamefont {I.}~\bibnamefont {Karpenko}}, \ and\
  \bibinfo {author} {\bibfnamefont {T.}~\bibnamefont {Pierog}},\ }\href@noop {}
  {\  (\bibinfo {year} {2013}{\natexlab{b}})},\ \Eprint
  {http://arxiv.org/abs/1312.1233} {arXiv:1312.1233 [nucl-th]} \BibitemShut
  {NoStop}%
%%CITATION = ARXIV:1312.1233;%%
\bibitem [{\citenamefont {Niemi}\ and\ \citenamefont
  {Denicol}(2014)}]{Niemi:2014wta}%
  \BibitemOpen
  \bibfield  {author} {\bibinfo {author} {\bibfnamefont {H.}~\bibnamefont
  {Niemi}}\ and\ \bibinfo {author} {\bibfnamefont {G.}~\bibnamefont
  {Denicol}},\ }\href@noop {} {\  (\bibinfo {year} {2014})},\ \Eprint
  {http://arxiv.org/abs/1404.7327} {arXiv:1404.7327 [nucl-th]} \BibitemShut
  {NoStop}%
%%CITATION = ARXIV:1404.7327;%%
\bibitem [{\citenamefont {Ortiz~Velasquez}\ \emph {et~al.}(2013)\citenamefont
  {Ortiz~Velasquez}, \citenamefont {Christiansen}, \citenamefont
  {Cuautle~Flores}, \citenamefont {Maldonado~Cervantes},\ and\ \citenamefont
  {Paic}}]{Ortiz:2013yxa}%
  \BibitemOpen
  \bibfield  {author} {\bibinfo {author} {\bibfnamefont {A.}~\bibnamefont
  {Ortiz~Velasquez}}, \bibinfo {author} {\bibfnamefont {P.}~\bibnamefont
  {Christiansen}}, \bibinfo {author} {\bibfnamefont {E.}~\bibnamefont
  {Cuautle~Flores}}, \bibinfo {author} {\bibfnamefont {I.}~\bibnamefont
  {Maldonado~Cervantes}}, \ and\ \bibinfo {author} {\bibfnamefont
  {G.}~\bibnamefont {Paic}},\ }\href {\doibase 10.1103/PhysRevLett.111.042001}
  {\bibfield  {journal} {\bibinfo  {journal} {Phys.Rev.Lett.}\ }\textbf
  {\bibinfo {volume} {111}},\ \bibinfo {pages} {042001} (\bibinfo {year}
  {2013})},\ \Eprint {http://arxiv.org/abs/1303.6326} {arXiv:1303.6326
  [hep-ph]} \BibitemShut {NoStop}%
%%CITATION = ARXIV:1303.6326;%%
\end{thebibliography}%

\end{document}